\documentclass[showpacs,preprintnumbers,12pt,onecolumn]{revtex4}
\usepackage{amsfonts}
\usepackage{amsmath}
\usepackage{graphicx}
\usepackage{dcolumn}
\usepackage{bm}

\input{tcilatex}

\begin{document}

\title{On the ratchet effect}
\author{Ricardo Chac\'{o}n$^{\ast }$}
\affiliation{Departamento de Electr\'{o}nica e Ingenier\'{\i}a Electromec\'{a}nica,
Escuela de Ingenier\'{\i}as Industriales, Universidad de Extremadura,
Apartado Postal 382, E-06071 Badajoz, Spain}
\author{Niurka R. Quintero}
\affiliation{Departamento de F\'{\i}sica Aplicada I, E. U. P., Universidad de Sevilla,
Virgen de \'{A}frica 7, 41011 Sevilla, Spain, and Instituto Carlos I de F%
\'{\i}sica Te\'{o}rica y Computacional, Universidad de Granada, 18071
Granada, Spain}
\date{\today}
\maketitle





\textbf{Understanding the ratchet effect induced by symmetry breaking of
temporal forces is a fundamental issue that has remained unresolved for
decades. While the dependence of the directed transport on each of the
ratchet-controlling parameters has been individually investigated
experimentally, theoretically, and numerically, there is still no general
criterion to apply to the whole set of these parameters to optimally control
directed transport in general systems. We report that, to optimally enhance
directed transport, there exists a universal force waveform which can be
understood as the result of two competing fundamental mechanisms: the
increase of the degree of breaking of the aforementioned temporal symmetries
and the decrease of the force impulse. We demonstrate that this universal
waveform explains all the previous experimental, theoretical, and numerical
results for a great diversity of systems, including motor enzymes.}

\textbf{\ }Consider a general system (classical or quantum, dissipative or
non-dissipative, one- or multi-dimensional, noisy or noiseless) where a
ratchet effect (\textit{1-3}) is induced by solely violating temporal
symmetries of a $T$-periodic zero-mean ac force $f\left( t\right) $ which
drives the system (\textit{4-18)}. This effect is important because of its
clear applicability to such diverse fields as biology, and micro- and
nano-technology (\textit{2,19-23}). In particular, it applies to electronic
transport through molecules (\textit{12}), smoothing surfaces (\textit{24}),
controlling vortex density in superconductors (\textit{22}), separating
particles (\textit{25,26}), controlling directed current in semiconductors (%
\textit{6,9}), and rectifying the phase across a SQUID (\textit{27}).\textbf{%
\ }A popular choice would be the simple case of a biharmonic force, 
\begin{equation}
f(t)=\epsilon _{1}har_{1}\left( \omega _{1}t+\varphi _{1}\right) +\epsilon
_{2}har_{2}\left( \omega _{2}t+\varphi _{2}\right) ,  \tag{1}
\end{equation}%
where $har_{1,2}$ represents indistinctly $\sin $ or $\cos $, and $p\omega
_{1}=q\omega _{2}$, $p,q$ coprime integers. In this case, the aforementioned
symmetries are the shift symmetry $\left( \mathbf{S}_{s}:f\left( t\right)
=-f\left( t+T/2\right) ,T\equiv 2\pi q/\omega _{1}=2\pi p/\omega _{2}\right) 
$ and the time-reversal symmetries $(\mathbf{S}_{tr,\pm }:$ $f(-t)=\pm f(t))$%
. Now a general unsolved problem is to find the regions of the parameter
space $\left( \epsilon _{i},\varphi _{i}\right) ,\epsilon _{1}+\epsilon
_{2}=const$., where the ratchet effect is optimal in the sense that the
average of relevant observables is maximal, the remaining parameters being
held constant. We show in the following that such regions are those where
the \textit{effective} degree of symmetry breaking is maximal. Without loss
of generality, we shall illustrate the degree of symmetry breaking (DSB)
mechanism by using the following working model for the driving force:%
\begin{eqnarray}
f_{ellip}(t) &=&\epsilon f(t;T,m,\theta )  \notag \\
&\equiv &\epsilon \limfunc{sn}\left( \Omega t+\Theta ;m\right) \limfunc{cn}%
\left( \Omega t+\Theta ;m\right) ,  \TCItag{2}
\end{eqnarray}%
where $\limfunc{cn}\left( \cdot ;m\right) $ and $\limfunc{sn}\left( \cdot
;m\right) $ are Jacobian elliptic functions of parameter $m$, $\Omega \equiv
2K(m)/T,$ $\Theta \equiv K(m)\theta /\pi $, $K(m)$ is the complete elliptic
integral of the first kind, $T$ is the period of the force, and $\theta $ is
the (normalized) initial phase $\left( \theta \in \left[ 0,2\pi \right]
\right) $. Fixing $\epsilon ,T$, and $\theta $, the force waveform changes
as the shape\textit{\ }parameter $m$ varies from 0 to 1 (see Fig. 1).
Physically, the motivation of the choice represented by Eq. 2 is that $%
f_{ellip}(t;T,m=0,\theta )=\epsilon \sin \left( 2\pi t/T+\theta \right) /2$,
and that $f_{ellip}(t;T,m=1,\theta )$ vanishes except on a set of instants
that has Lebesgue measure zero, i.e., in these two limits directed transport
is not possible, while it is expected for $0<m<1$. Thus, one may expect 
\textit{in general} the average of any relevant observable $\Re $ to exhibit
an extremum at a certain critical value $m=m_{c}$ as the shape parameter $m$
is varied, the remaining parameters being held constant. In this work, we
demonstrate that such a value $m_{c}$ is universal, i.e., there exists a
universal force waveform which optimally enhances the ratchet effect in any
system. Clearly, there are two competing fundamental mechanisms which allow
one to understand the appearance of such extremum: the increase of the
degree of breaking of the shift symmetry as $m$ is increased, which
increases the absolute value of the average, and the effective narrowing of
the force pulse as $m$ is increased, which decreases the absolute value of
the average. The former mechanism arises from the fact that a broken
symmetry is a structurally stable situation (Curie's principle) and can be
quantitatively characterized by noting that%
\begin{equation}
\frac{-f\left( t+T/2\right) }{f\left( t\right) }=\frac{\sqrt{1-m}}{\limfunc{%
dn}^{2}\left( \Omega t+\Theta ;m\right) }\equiv D\left( t;T,m,\theta \right)
,  \tag{3}
\end{equation}%
where $\limfunc{dn}\left( \cdot ;m\right) $ is the Jacobian elliptic
function. Equation 3 indicates that the degree of deviation from the shift
symmetry condition $\left( D\left( t;T,m,\theta \right) \equiv 1\right) $
increases on average as $m\rightarrow 1$, irrespective of the values of the 
\textit{period} and \textit{initial phase} (see Fig. 2). Thus, while
increasing the shape parameter from $m$ $\left( 0<m<m_{c}\right) $ improves
the directed transport yielding a higher average, it simultaneously narrows
the force pulse lowering the driving effectiveness of the force. Indeed, the
latter effect becomes dominant for sufficiently narrow pulses $\left(
m>m_{c}\right) $. Next, one exploits the universality of the waveform
corresponding to $m_{c}$ to deduce the optimal values of the parameters of
the biharmonic force of Eq. 1. To this end, notice that we chose the
function of Eq. 2 to satisfy the requirement that $m_{c}$ be sufficiently
far from 1 so that the elliptic force is effectively approximated by its
first two harmonics. One thus obtains a relationship between the amplitudes
of the two harmonics in parametric form: $\epsilon _{1,2}=\epsilon _{1,2}(m)$%
. This relationship does not depend on the initial phase $\theta $, and
hence neither does it depend on the breaking of time-reversal symmetries of
the biharmonic approximation corresponding to the elliptic force. For a
general biharmonic force (Eq. 1), this means according to the DSB mechanism
that the optimal ratchet-inducing values of the initial phases $\varphi
_{1},\varphi _{2}$ should be those giving a maximal breaking of one of the
two time-reversal symmetries of the force, while the relationship $\epsilon
_{2}=\epsilon _{2}(\epsilon _{1};p,q)$ should control solely the degree of
breaking of the shift symmetry. Note that this symmetry is \textit{not}
broken when $p,q$ are both odd integers. Consequently, if the DSB mechanism
is right, the relationship $\epsilon _{2}=\epsilon _{2}(\epsilon _{1};p,q)$ (%
$p+q=2n+1,n=1,2,...$) controlling the degree of breaking of the shift
symmetry should be independent of the particular system in which directed
transport is induced. This implies that any averaged observable $<\Re >$
should be proportional to a certain function $g\left( \epsilon _{1},\epsilon
_{2}\right) \equiv g\left( \epsilon _{1},\epsilon _{2};p,q\right) $ which is 
$\sim p_{1}\left( \epsilon _{1}\right) p_{2}\left( \epsilon _{2}\right) $ in
leading order with $p_{1}(\epsilon _{1})\sim \epsilon
_{1}^{r},p_{2}(\epsilon _{2})\sim \epsilon _{2}^{s}$, $r,s$ positive
integers. Since the aforementioned extremum $m_{c}$ is independent of the
driving amplitude, one defines $\epsilon _{1}=\epsilon \left( 1-\alpha
\right) ,\epsilon _{2}=\epsilon \alpha $ $\left( \alpha \in \left[ 0,1\right]
\right) $, so that $g\left( \epsilon _{1},\epsilon _{2}\right) \sim \left(
1-\alpha \right) ^{r}\alpha ^{s}$ taking $\epsilon =1$ without loss of
generality. Since the extremum $m_{c}$ is also independent of the driving
period, one has the symmetry relationship $g\left( \epsilon _{1},\epsilon
_{2};p,q\right) =g\left( \epsilon _{2},\epsilon _{1};q,p\right) $. The
problem thus reduces to finding the relationship between $\left( r,s\right) $
and $\left( p,q\right) $. From Maclaurin's series, one sees that the \textit{%
simplest} function satisfying all these requirements in leading order is $%
\left( 1-\alpha \right) ^{p}\alpha ^{q}$, and hence $g\left( \epsilon
_{1},\epsilon _{2};p,q\right) \sim $ $\epsilon _{1}^{p}\epsilon _{2}^{q}$.
Indeed, previous theoretical analyses of every kind on a great diversity of
systems (\textit{5-11,14,15,17,18}) have found that the averaged observable
is always proportional to such a factor in leading order. In particular, one
thus obtains%
\begin{equation}
<\Re >\sim \epsilon ^{3}S(m)\equiv \epsilon ^{3}\frac{\func{sech}^{2}\left[ 
\frac{\pi K(1-m)}{K(m)}\right] \func{sech}\left[ \frac{2\pi K(1-m)}{K(m)}%
\right] }{m^{3}K^{6}(m)},  \tag{4}
\end{equation}%
for the biharmonic approximation corresponding to the elliptic force (i.e., $%
p=2,q=1$). Therefore, the shape function $S(m)$ is a universal function
which controls the breaking of the shift symmetry in leading order for the
resonance $(p,q)=(2,1)$. It presents a single maximum at $m=0.960057\simeq
m_{c}$ for which $\epsilon _{2}=\epsilon _{2}(\epsilon _{1})\simeq
0.396504\epsilon _{1}$ (note that $\epsilon _{2}=\epsilon _{1}/2$ for $%
m=0.983417$; see Fig. 3). Since the DSB mechanism is \textit{%
scale-independent}, the critical value $m_{c}$ could well be defined by a
purely geometric condition (Occam's razor) which takes into account the two
aforementioned competing mechanisms (degree of breaking of symmetries and
transmitted impulse): $A(m=m_{c})/A(m=0)=\Phi /2$, where $A(m)\equiv
\int_{0}^{T/2}f_{ellip}(t;T,m,\theta =0)dt$ and $\Phi =\left( \sqrt{5}%
+1\right) /2$ is the golden ratio. This gives $m_{c}=0.9830783...$. The
waveform corresponding to this value can be very accurately approximated by
a sawtooth of unit height $\left( h=1\right) $, unit period $\left( \lambda
=\lambda _{1}+\lambda _{2}=1\right) $, critical asymmetry parameter $%
a_{c}\equiv \lambda _{1}/\lambda =1/4$, and critical tangents $\tan \theta
_{c,1}\equiv h/\lambda _{1}=4$, $\tan \theta _{c,2}=h/\lambda _{2}=4/3$ with 
$\theta _{c,2}\equiv 4\arctan \Phi $.

We now discuss the application of the DSB mechanism to time-reversal
symmetries. For the sake of clarity, consider first the case with no
dissipation. According to the above discussion, the optimal ratchet-inducing
values of the initial phases $\varphi _{1},\varphi _{2}$ should be those
yielding a maximal interference-type effect of the two harmonic functions in
Eq. 1, i.e., \textit{in general}, $\left\vert q\varphi _{2}-p\varphi
_{1}\right\vert _{c}$ can solely adopt some of the values $\left\{ 0,\pi
/2,\pi ,3\pi /2\right\} $ depending upon the particular harmonic functions $%
har_{1,2}$ and the particular values $p,q$. When dissipation is not
negligible, two effects are expected: The average $<\Re >$ will decrease as
dissipation is increased, and the critical values of the initial phases will
be shifted from their values at the limiting dissipationless case ($\varphi
_{c,i}\rightarrow \varphi _{c,i}+\varphi _{0})$. Note that the DBS mechanism
implies that the dissipation phase $\varphi _{0}\rightarrow 0$ as
dissipation vanishes.

We found that the present theory confirms and explains all the previous
experimental, theoretical, and numerical results for a great diversity of
systems subjected to a biharmonic force (Eq. 1) (\textit{5-11,14,15,17,18}).
In particular, it explains recent experimental results of directed diffusion
in a symmetric optical lattice (\textit{13}), where the force $F=(\limfunc{%
constant})\left[ \left( 1-B\right) \cos \left( \omega t\right) +B\cos \left(
2\omega t-\phi \right) \right] $ yielded the maximum velocity of the
centre-of-mass of the atomic cloud at $B\simeq 0.33$, $\phi =\pi /2$, and
the complete data series fitted the functional form $(1-B)^{2}B$ (i.e., $%
\epsilon _{1}^{2}\epsilon _{2}$), in confirmation of the predictions above.
The analysis of the problem of electron tunneling through a one-dimensional
potential barrier in the presence of an externally applied ac force $%
f_{1}\sin \left( \omega t\right) +f_{2}\sin \left( 2\omega t+\varphi \right) 
$ (\textit{11}) showed that the directed current is proportional to $%
2Af_{1}^{2}f_{2}\cos \varphi $, in confirmation of the present theory. Also,
the DSB mechanism works powerfully in the case of electron transport through
a molecular wire weakly coupled to two leads and subjected to a biharmonic $%
\left( A_{1}\sin \left( \Omega t\right) +A_{2}\sin \left( 2\Omega t+\phi
\right) \right) $ laser field (\textit{12}). Quantum calculations showed (%
\textit{12}) that only when the symmetry breaking is maximal $\left( \text{%
i.e., }A_{2}=A_{1}/2,\phi =\pi /2\right) $ does the average current reach
its maximum value, and that the average current is proportional to the
coupling strength. For weak coupling, this maximum is about five orders of
magnitude higher than the corresponding value at $\phi =0$. We also found
that the universality of the present theory is confirmed by previous studies
on systems subjected to sawtooth-shaped excitations (\textit{26,28). }%
Indeed, a study of horizontal transport of granular layers on a vertically
vibrated sawtooth-shaped base (\textit{26}) showed that the horizontal
transport $\left\langle v_{x}\right\rangle $ versus asymmetry parameter $a$
has an extremum at $a=0.25$ only when the sawtooth waveform approximated the
above universal waveform (i.e., $\lambda =6$ mm, $h=5$ mm, cf. (\textit{26}%
); note that the universal waveform is recovered for $h=6$ mm). Although we
have here limited ourselves to temporal symmetries, the DSB mechanism
applies to spatial symmetries as well. In this regard, one should expect
that the ratchet potential underlying biological motor proteins might be
optimized according to the DSB mechanism. A clear experimental confirmation
of this prediction appears in the context of the actomyosin dynamics in the
presence of external loads (\textit{28}), where the mean first-exit time
versus external applied force was studied for a sawtooth potential profile
with different values of the asymmetry parameter. The experimental data were
optimally fitted only when the sawtooth waveform approximated the above
universal waveform (i.e., $\tan \theta _{1}\equiv h/\lambda _{1}\simeq 4.55$%
, cf. (\textit{28}); recall that $\tan \theta _{c,1}=4$ for the universal
waveform).

We here studied the example of kink-asisted directed energy transport in a
driven, damped, sine-Gordon (sG) equation (see Fig. 4), where directed
energy transport is predicted for a non-zero topological charge, implying
the existence of sG solitons (kinks) in the system. As Fig. 5 shows, the
theoretical predictions are confirmed by numerical simulations even in the
presence of noise. We should stress that the consequences of the present
theory extend beyond the problem of directed transport. It applies, for
example, to the phenomenon of synchronization of arrays of coupled
limit-cycle oscillators (\textit{29}), where the maximal symmetry breaking
of a homogeneous, time-delayed, periodic coupling gave the maximum decrease
of the synchronization frequency (see Fig. 6). Also, the DBS mechanisms
explains the effectiveness of harmonic mixing signals (with frequencies $%
\omega ,2\omega $ and amplitudes $\eta _{\omega },\eta _{2\omega }$,
respectively) in controlling (enhancing and suppressing) stochastic
resonance phenomena (\textit{30,31}). In particular, it explains the
dependence of the output power at the signal frequency $\omega $ ($P_{2}\sim
\eta _{\omega }^{2}\eta _{2\omega }$, cf. (\textit{30})) in a modified
Schmitt trigger electronic circuit, and the dependence of the time- and
noise-averaged mean value of the response ($\gamma _{0}\sim $ $\eta _{\omega
}^{2}\eta _{2\omega }$, cf. (\textit{31})) of an overdamped two-well Duffing
oscillator. We should stress that the present theory can be directly applied
to optimizing the effectiveness of a two-colour laser field for strong
harmonic generation [32] as well as to optimally designing synthetic
molecular motors [33].

* Author to whom all correspondence should be addressed.

{\Large Figure Captions}

Figure 1. Function $f(t;T,m,\theta )$ (Eq. 2) vs $t/T$ for $\theta =0$ and
three shape parameter values, $m=0,1-10^{-6}$ (cyan), and $0.96$ (magenta),
showing an increasing symmetry-breaking sequence as the pulse narrows, i.e.,
as $m\rightarrow 1$.

Figure 2. Deviation from the shift symmetry condition $D\left( t;T,m,\theta
=0\right) =1$ (Eq. 3) showing an increasing deviation as $m\rightarrow 1$.

Figure 3. Universal shape function $S(m)$ in leading order for the resonance 
$(p,q)=(2,1)$ (Eq. 4) exhibiting a maximum at $m=0.960057$.

Figure 4. Average velocity of the kink centre-of-mass versus shape parameter
for the sG equation $U_{tt}-U_{xx}+\sin \left( U\right) =-\beta
U_{t}+f_{ellip}(t)$ with $\epsilon =0.1,\beta =0.05,T=20\pi ,$ and $\theta
=0 $. Previous results (\textit{14}) from a collective coordinate approach
with two degrees of freedom, $X(t)$ and $l(t)$ (respectively, position and
width of the kink), can be directly applied to obtain an ODE governing the
dynamics of these two collective coordinates in the presence of the elliptic
force (Eq. 2): $\overset{.}{P}=-\beta P-qf_{ellip}(t)$, $\overset{..}{l}=%
\overset{.}{l}^{2}/(2l)+1/(2\alpha l)-\beta \overset{.}{l}-(\Omega
_{R}^{2}l/2)(1+M_{0}^{-2}P^{2})$, where the momentum $P(t)=M_{0}l_{0}\overset%
{.}{X}/l(t)$, $\Omega _{R}=\sqrt{12}/(\pi l_{0})$ is the Rice frequency, $%
\alpha =\pi ^{2}/12$, and $M_{0}=8$, $q=2\pi $, and $l_{0}=1$ are,
respectively, the dimensionless kink mass, topological charge, and
unperturbed width. For the biharmonic approximation corresponding to the
elliptic force (2), one straightforwardly obtains the following estimate for
the average velocity of the kink: $\left\langle \overset{.}{X}\left(
t\right) \right\rangle =\epsilon ^{3}F(\beta ,T)S\left( m\right) $, where $%
S(m)$ is the shape function (Eq. 4) and $F(\beta ,T)$ provides the
dependence upon the dissipation and the period (\textit{14}). The solid,
long-dashed, and dot-dashed lines represent, respectively, the results from
the numerical solution of the sG PDE, the results from the numerical
solution of the collective coordinates ODE, and the above analytical
estimate $\left\langle \overset{.}{X}\left( t\right) \right\rangle $. The
curves corresponding to the two kinds of numerical solution present minima
at $m=0.983417$ (solid line; recall that $m_{c}=0.9830783$) and $m=0.97904$
(long-dashed line), respectively, while the curve corresponding to the
analytical approximation has its minimum at $m=0.960057$.

Figure 5. Average velocity of the kink centre-of-mass versus shape parameter
for the sG equation $U_{tt}-U_{xx}+\sin \left( U\right) =-\beta
U_{t}+f_{ellip}(t)+\sqrt{D}\eta \left( x,t\right) $ with $\epsilon
=0.1,\beta =0.05,T=20\pi ,D=0.001$ and $\theta =0$, where $\eta (x,t)$ is a
Gaussian white noise of zero mean and correlations $\left\langle \eta \left(
x,t\right) \eta \left( x^{\prime },t^{\prime }\right) \right\rangle =\delta
\left( x-x^{\prime }\right) \delta \left( t-t^{\prime }\right) $. The lines
represent the noiseless cases described in Fig. 4, while circles represent
the results from the numerical solution of the noisy sG PDE which presents a
minimum at $m=0.986525$ (recall that $m_{c}=0.9830783$).

Figure 6. Symmetry-breaking-induced frequency suppression versus shape
parameter for the system of $N$ coupled oscillators $d\phi _{i}(t)/dt=\omega
_{0}+2\epsilon \sum_{j}f_{ellip}\left[ \phi _{j}\left( t-\tau \right) -\phi
_{i}\left( t\right) ;T,m,\theta \right] $ with $T=2\pi ,\theta =0$, and
where $\epsilon $ is the coupling constant, $\tau $ is the delay, and $%
\omega _{0}$ is the intrinsic frequency of the oscillators. The lowest
stable frequency associated with the synchronization states $\left( \phi
_{i}=\Omega t+\Omega _{0}\right) $ for the biharmonic approximation
corresponding to the elliptic coupling is given by $\Omega _{\min }\approx
\omega _{0}/[1+2\pi ^{2}n\tau \epsilon g(m)]$, where $n$ is the number of
neighbours and $g(m)\equiv \frac{1}{mK^{2}(m)}\left\{ \func{sech}\left[ 
\frac{\pi K(1-m)}{K(m)}\right] +4\func{sech}\left[ \frac{2\pi K(1-m)}{K(m)}%
\right] \right\} $. The curve exhibits a minimum at $m=0.9845$ (recall that $%
m_{c}=0.9830783$) for $n=4,\tau =0.1$, and $\epsilon =3$. Note that this
minimum is the same for all values of $\tau ,n$, and $\epsilon $, in
confirmation of the present theory.

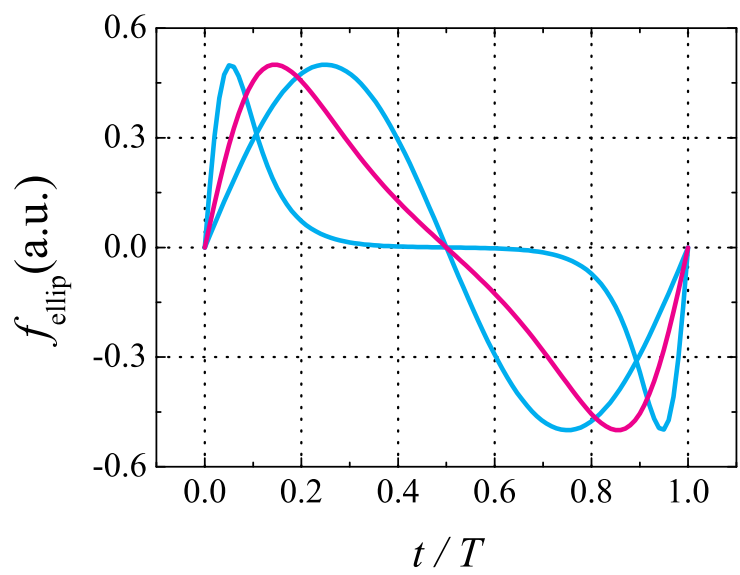
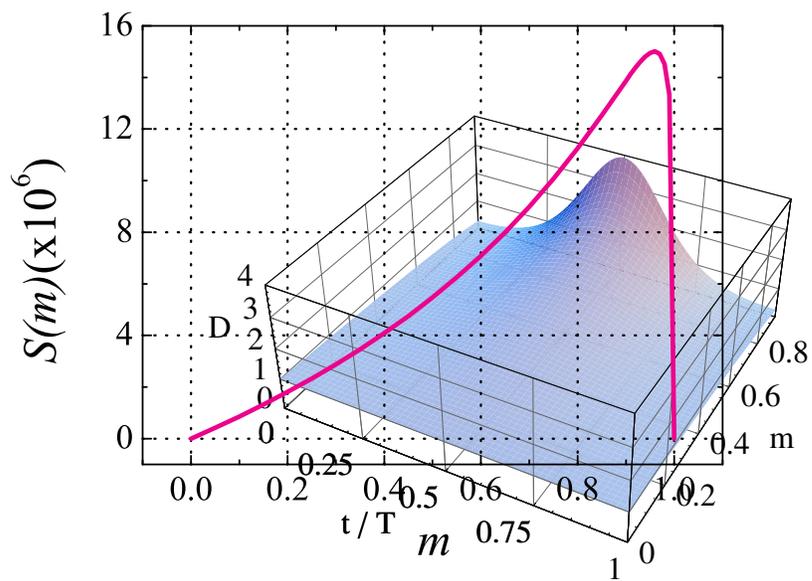
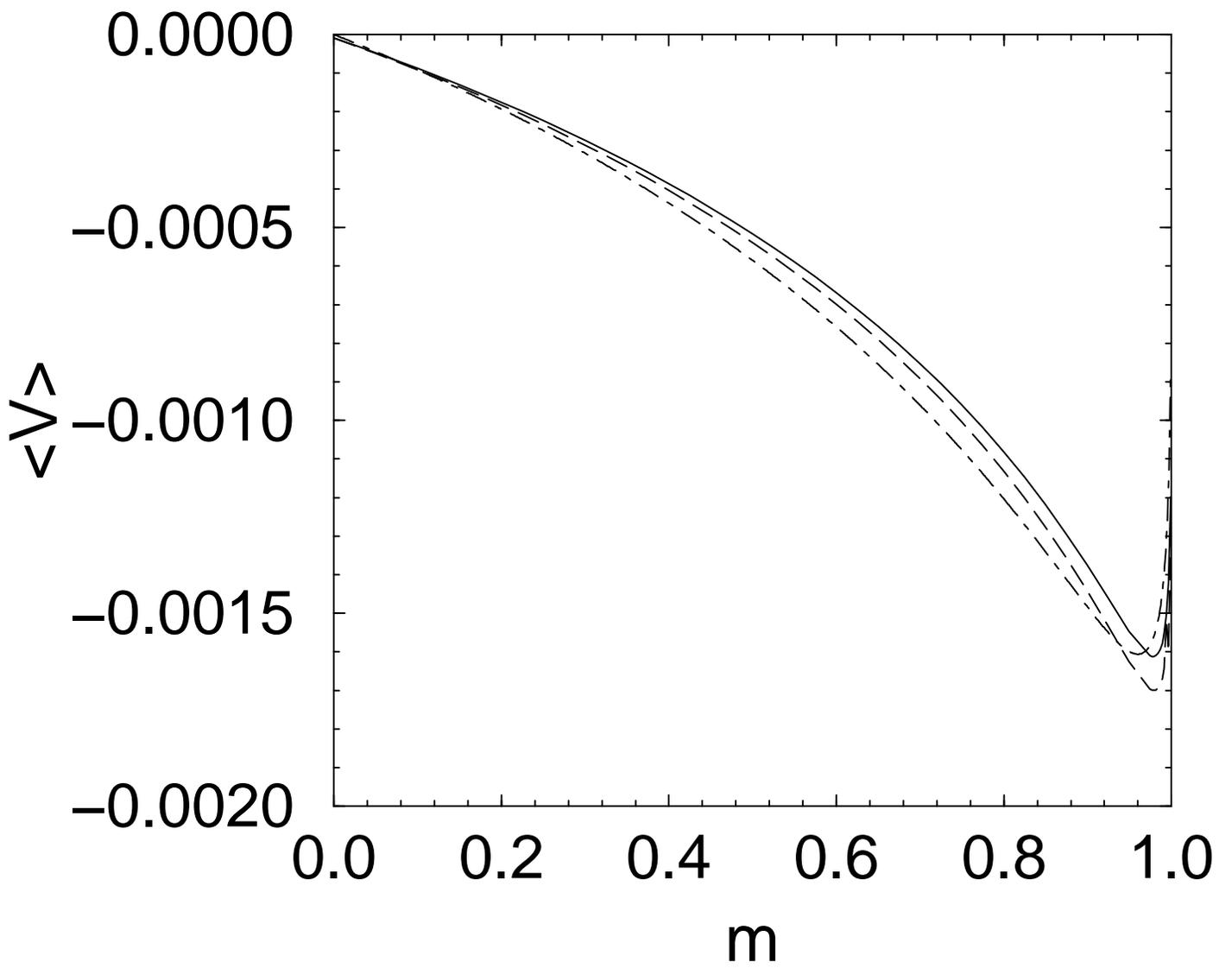
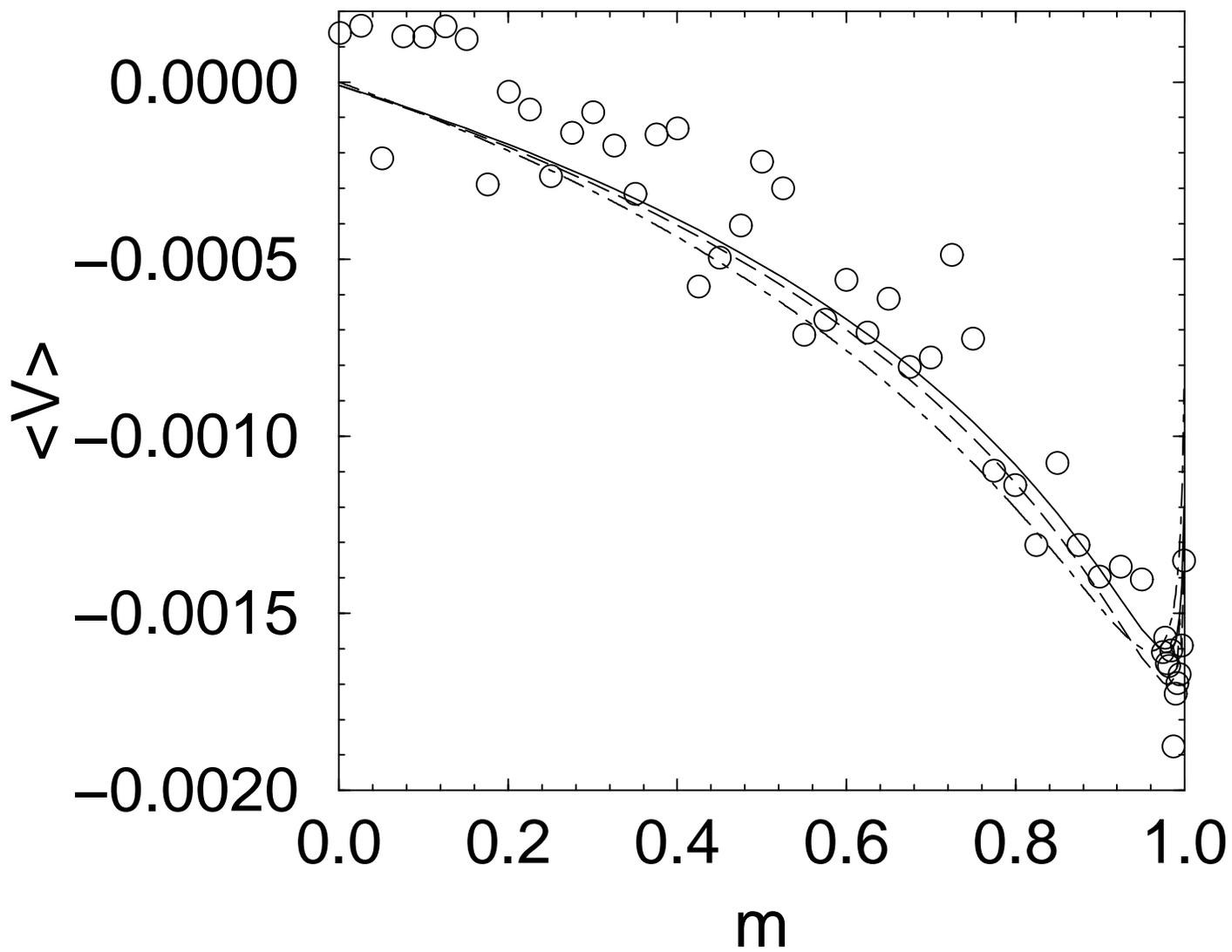
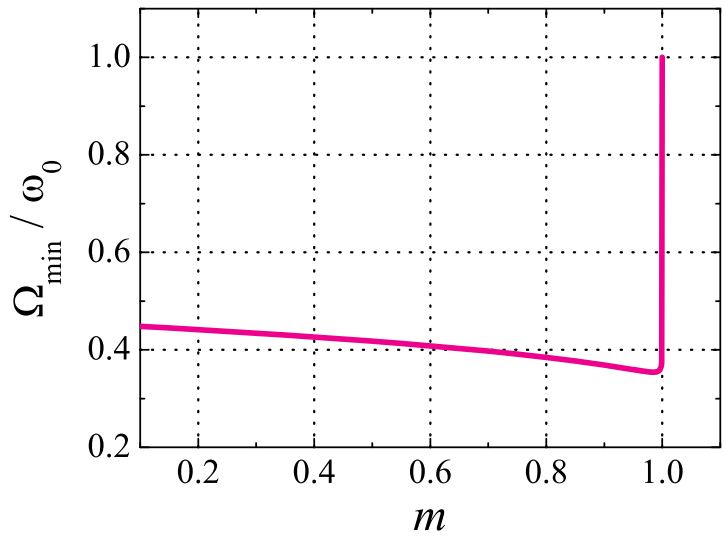
\end{document}